\documentclass{iaus}
\usepackage{graphicx}

\title[Instabilities: from microquasars to quasars] %% give here short title %%
{Probing the accretion disk - jet connection via instabilities in the inner accretion flow. From microquasars to quasars}

\author[Agnieszka Janiuk et al.]   %% give here short author list %%
{Agnieszka Janiuk$^1$, Bo\.zena Czerny$^2$, Monika Mo\'scibrodzka$^3$,
 \and Aneta Siemiginowska$^4$}

\affiliation{$^1$Center for Theoretical Physics, Polish Academy of Sciences, Al. Lotnikow 32/46, 02-668 Warsaw, Poland
$^2$N. Copernicus Astronomical Center, Bartycka 18, 00-716
Warsaw, Poland \\[\affilskip]
$^3$ Department of Physics, University of Illinois, 1110 West Green Street, Urbana, IL 61801, USA \\[\affilskip]
$^4$ Harvard Smithsonian Center for Astrophysics, 60 Garden St, 
Cambridge, MA 02138, USA\\[\affilskip]
}

\pubyear{2010}
\volume{275}  %% insert here IAU Symposium No.
\pagerange{1-1}
\jname{Jets at all scales}
\editors{Gustavo E. Romero, Rashid A. Sunyaev \& Tomaso Belloni}
\begin{document}

\maketitle

\begin{abstract}

We present various instability mechanisms in the accreting black hole 
systems which might indicate at the connection between the
accretion disk and jet. The jets observed in microquasars
can have a peristent or blobby morphology. Correlated with the
accretion luminosity, this might provide a link to the cyclic outbursts
of the disk. Such duty-cycle type of behaviour on short timescales
results from the thermal instability caused by the radiation pressure domination.
The same type of instability may explain the cyclic radioactivity of the
supermassive black hole systems.  The somewhat longer timescales
are characteristic for the instability 
caused by the partial hydrogen ionization.
The distortions of the jet direction and complex morphology
of the sources can be caused by precesion of the disk-jet axis. 
%\keywords{Keyword1, keyword2, keyword3, etc.}
\end{abstract}

%\firstsection 

\section{Overview}

\begin{figure}[b]
% \vspace*{-2.0 cm}
%\begin{center}
 \includegraphics[width=2.0in]{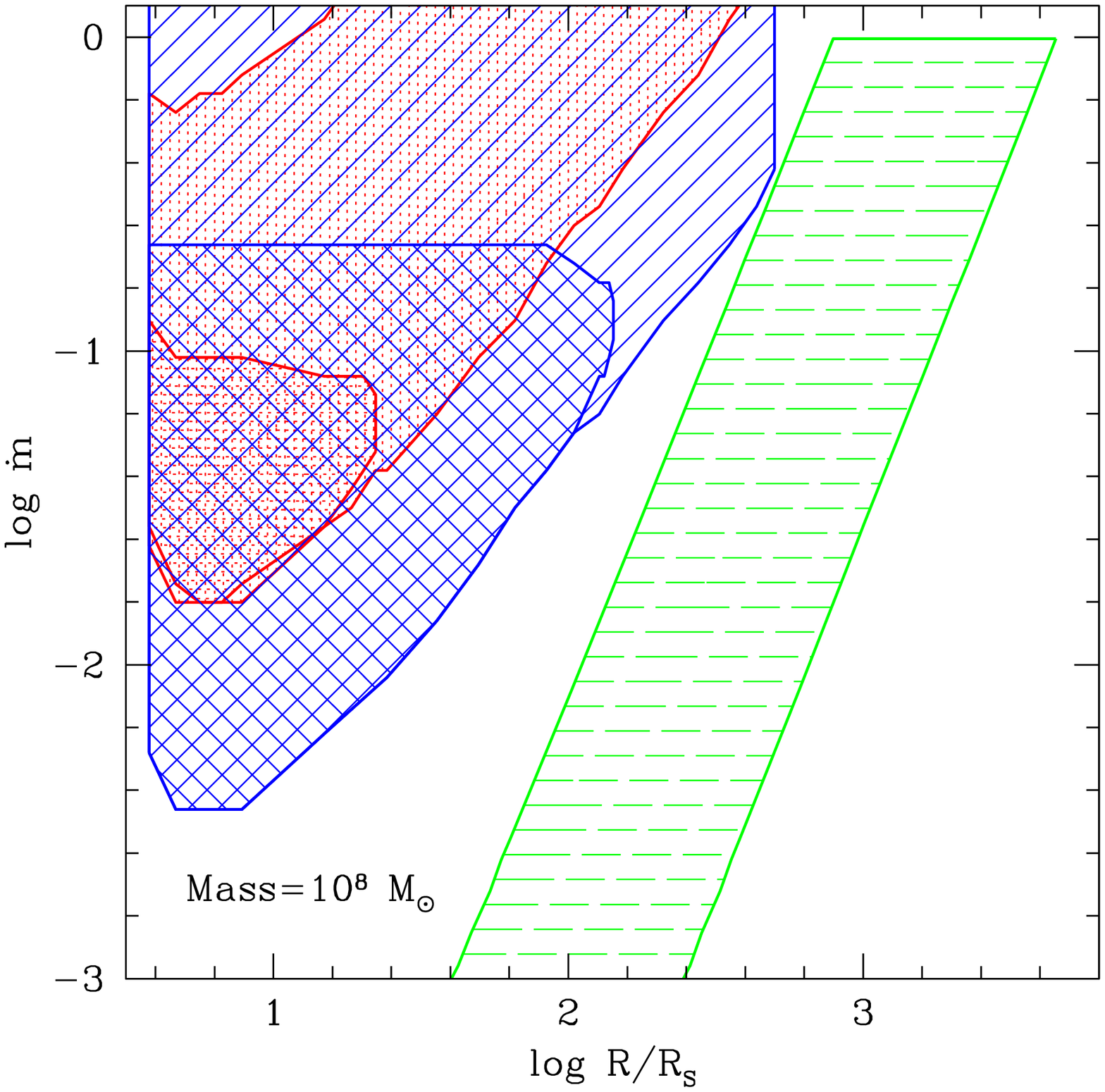} 
 \includegraphics[width=2.0in]{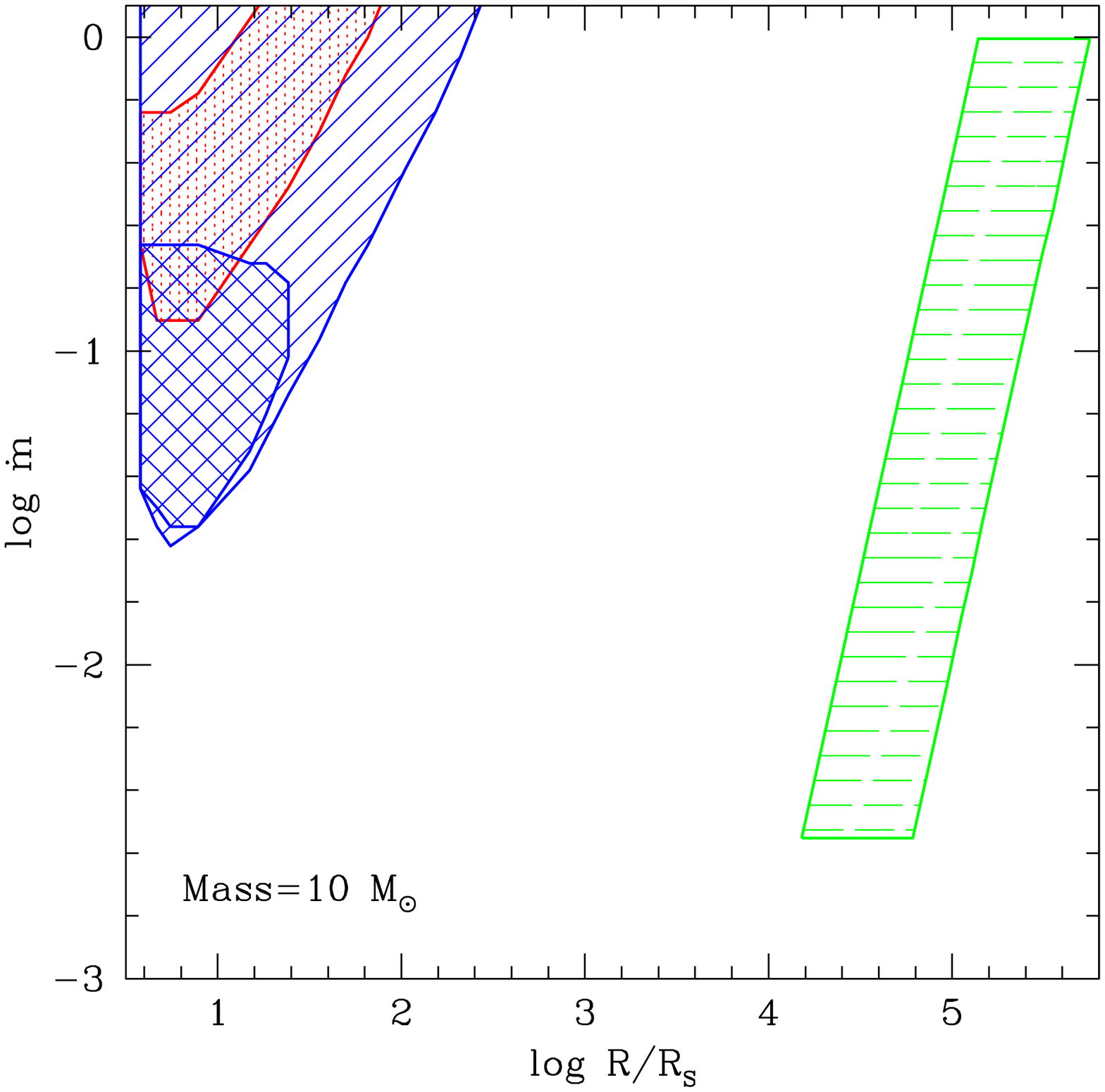} 
% \vspace*{-1.0 cm}
 \caption{Extension of the radiation pressure (solid and dotted lines)
and hydrogen ionization (dashed lines) unstable zones, depending on mean accretion
rates (Eddington units). The results are for two heating prescriptions, as well
as a zero and non-zero fraction of jet power.}
   \label{fig:fig1}
%\end{center}
\end{figure}

{\underline{\it Radiation pressure instability}}.
The black hole accretion disk is
subject to the thermal and viscous instability if the radiation
pressure dominates over the gas pressure.  
If the accretion rate outside the unstable region
(i.e. the mean accretion rate) is such that the disk is
unstable, the source enters a cycle of bright, hot states,
separated by the cold, low luminosity states. 
The outburst amplitudes and durations
are sensitive to black hole mass, viscosity and the
mean accretion rate.  
If the viscous heating is proportional to the total pressure, the outburst
amplitudes are very large - they can be reduced if the heating is 
proportional to the square
root of the gas times the total pressure. 

{\underline{\it Partial Hydrogen ionization instability}}.
Another type of the 
thermal - viscous instability is due to the partial
hydrogen ionization. The disk also 
cycles between the two states: a
hot and mostly ionized state with a large local accretion rate and a
cold, neutral state with a low accretion rate. This instability has
originally been proposed to explain the luminosity
variations observed in cataclysmic variables .
The same mechanism
is responsible for the eruptions in soft X-ray transients (see
e.g., \cite{Lasota01} for review) as well as may
operate in active galactic nuclei (e.g., \cite{jan04}).

{\underline{\it Disk precession}}. As studied
by \cite{jan09}, the acoustic instabilities of the
Papaloizou-Pringle type can arise in the supersonic inner parts
of the accretion flow. The azimuthal modes of such instabilities will
lead to the tilt of the disk and its precession.

\section{Implications}

We speculate that in the hot state, the
luminous core will power a radio jet,
while during the cold
state the radio activity ceases.
%{\bf a bit more discussion ???}
In case of the regular periodic outbursts of
GRS~1915+105 (see e.g. \cite{fb04}), 
lasting from $\sim 100$ to $\sim 2000$ s, 
%(depending on the source mean luminosity), 
this approach is successful (e.g.
\cite{jan02}).
%No other quantitative mechanism has been proposed to explain the observed 
%behavior of this object, and
Only the limit cycle mechanism (likely driven by the radiation pressure
instability) explains the absence of the direct transitions from its
state C to the state B, 
as well as the observed QPO
oscillations. 
Scaling the timescale with the black hole mass by a factor $10^8$ gives
the outbursts durations of $10^{2} - 10^{4}$ yrs, and amplitudes are
sensitive to the energy fraction deposited in the jet (see Fig. \ref{fig:fig1}).
%This gives an 
%additional, model-independent argument that an intermittency in AGN on 
%the timescales of hundreds/thousands of years is likely of a similar origin. 
The radiation pressure 
instability operating above $\dot m=0.025$
may be responsible for very short ages of Compact Symmetric Objects
(\cite {bcz09}).
%), a subgroup of GigaHertz Peaked Sources (GPS).
The radiostructures may be influenced by jet precession
(\cite{kun10}).

In the ionization instability, the derived separations
between outbursts are on order of $10^6$\,yrs for a $10^8\,M_{\odot}$
black hole, while the outburst duration is an order of magnitude
shorter. The
location of the unstable zone is sensitive to the black hole mass (see Fig. \ref{fig:fig1}).
For an X-ray binary, the disk maximum radius is limited by the Roche-lobe 
size and might be too small for an appropriate temperature range to appear.
In AGN, the partly ionized zone is much closer to the black hole.
If the two unstable zones are very close to each other, the
rate of supply of material to the radiation pressure dominated region may be modulated
on long timescales, independently of the environment in the host galaxy.
%This will influence the overally very complex pattern of evolution of AGN cores.

{\bf Acknowledgments} This work was supported in part by grant NN 203 512638 
from the Polish Ministry of Science.


\begin{thebibliography}{}

\bibitem[Czerny et al. (2009)]{bcz09} {Czerny, B., et al} 2009, \textit{ApJ}, 698, 840 
\bibitem[Fender \& Belloni (2004)]{fb04} {Fender, R., Belloni, T.} 2004, \textit{ARAA}, 42, 317
\bibitem[Janiuk et al. (2002)]{jan02} {Janiuk, A., Czerny, B., \& Siemiginowska, A.} 2002,\textit{ ApJ}, 576, 908
\bibitem[Janiuk et al. (2004)]{jan04} {Janiuk, A., Czerny, B., Siemiginowska, A., \& Szczerba, R.,} 2004,\textit{ApJ}, 602, 595
\bibitem[Janiuk et al.(2009)]{jan09} {Janiuk, A., Sznajder, M., Moscibrodzka,
M., Proga, D.} 2009, \textit{ApJ}, 705, 1503
\bibitem[Kunert-Bajraszewska et al.(2010)]{kun10} {Kunert-Bajraszewska, M., Janiuk, A., Gawronski, M., Siemiginowska, A.} 2010, \textit{ApJ}, 718, 1345
\bibitem[Lasota (2001)]{Lasota01} {Lasota, J.-P.}, 2001, \textit{New Astron. Revs.}, 45, 449 

\end{thebibliography}
\end{document}